\documentclass[aps,prl,preprint,groupedaddress]{revtex4-1}

 \usepackage{graphics,graphicx,subfigure,bm}
    \usepackage{amsmath,amssymb,pifont}
    \usepackage{caption}
    \usepackage[section]{placeins}
    \usepackage{natbib}
    \usepackage[usenames,dvipsnames]{color}
\usepackage{graphicx,color,xfrac}

\graphicspath{{Figures/}}

\def \bn {\mathbf{n}}
\def \d {\mathrm{d}}
\def \De {\mathrm{De}}

\def \bu {\mathbf{u}}
\def \taup {\boldsymbol{\tau}_p}

\def \bsig {\boldsymbol{\sigma}}

\def \bA {\mathbf{A}}

\def \bez {\mathbf{e}_z}

\def \strain {\mathbf{E}}

\def \Des {\De_{\mathrm{so}}}

\def\varepsilon{\epsilon}
\def\De{{\text{De}}}

\def \lp  {\left(}
\def \rp  {\right)}

\def \bA   {\mathbf{A}}

\def \nnabla {\tilde{\boldsymbol{\nabla}}}

\def \hbar {\bar{h}}

\def \bsig {\boldsymbol{\sigma}}

\def \bI {\mathbf{I}}

\usepackage{color}
\usepackage{tikz}
\usetikzlibrary{shapes}
\newcommand{\markerDone}{\raisebox{0pt}{\tikz{\node[draw=green,scale=0.6,regular polygon, regular polygon sides=4,fill=green](){};}}}
\newcommand{\markerDtwo}{\raisebox{0pt}{\tikz{\node[draw,scale=0.6,circle,fill=black!20!black](){};}}}
\newcommand{\markerDthree}{\raisebox{0pt}{\tikz{\node[draw=blue,scale=0.5,diamond,fill=blue!10!blue](){};}}}
\newcommand{\markerDfour}{\raisebox{0pt}{\tikz{\node[draw=red,scale=0.4,regular polygon, regular polygon sides=3,fill=red!45!red,rotate=180](){};}}}
\newcommand{\markerDfive}{\raisebox{0pt}{\tikz{\node[draw=cyan,scale=0.4,regular polygon, regular polygon sides=3,fill=cyan](){};}}}
\newcommand{\markerDsix}{\raisebox{0pt}{\tikz{\node[draw,scale=0.6,circle,fill=magenta](){};}}}

\begin{document}


\title{Viscoelastic levitation}

\author{Yunxing Su$^{1}$, Alfonso Castillo$^{2,3}$, On Shun Pak$^ {4}$, Lailai Zhu$^{5}$ and Roberto Zenit$^{1,2}$ }

\affiliation{
$^{1}$Center for Fluid Mechanics, School of Engineering, Brown  University, Providence, RI 02912, USA;
$^{2}$Instituto de Investigaciones en Materiales, Universidad Nacional Aut\'onoma de M\'exico, Ciudad de M\'exico, 04510, M\'exico;
$^{3}$Departamento de Ingenier\'ia Qu\'imica, Facultad de Qu\'imica, Universidad Nacional Aut\'onoma de M\'exico, Ciudad de M\'exico, 04510, M\'exico;
$^{4}$Department of Mechanical Engineering, Santa Clara University, Santa Clara, California, 95053, USA;
$^{5}$Department of Mechanical Engineering, National University of Singapore, 117575, Singapore
}



\date{\today}

\begin{abstract}
The effects of viscoelasticity have been shown to manifest themselves via symmetry breaking. In this investigation, we show a novel phenomenon that arises from this idea. We observe that when a dense sphere is rotated near a wall (the rotation being aligned with the wall-normal direction and gravity), it levitates to a fixed distance away from the wall. Since the shear is larger in the gap (between the sphere and the wall) than in the open side of the sphere, the shear-induced elastic stresses are thus asymmetric, resulting in a net elastic vertical force that balances the weight of the sphere.
We conduct experiments, theoretic models, and numerical simulations for rotating spheres of various sizes and densities in a Boger-type fluid. In the small Deborah number range, the results are  collapsed into a universal trend by considering a dimensionless group of the ratio of elastic to gravitational forces.
\end{abstract}

\maketitle

\section{Introduction}

The study of classical Newtonian fluid flows constitutes the foundation of fluid mechanics. Through experiments, theory and numerical solutions, we have gained a vast insight of the nature of flow for a wide range of conditions, from laminar to turbulent. The situation is very different for complex fluids \citep{Larson:1999}. In many such cases, the presence of memory and stress anisotropy substantially change the nature of the flow, leading to dramatic differences. For instance, a two-dimensional shear flow gives rise to non-zero normal stresses in a viscoelastic fluid, unlike in a Newtonian fluid. Many of the surprising phenomena seen in the flow of complex fluids, and in viscoelastic fluids in particular, can be understood by an examination of these normal stresses and the normal stress differences \citep{Morozov2015}.

The general mechanism for the appearance of normal stress can be explained by the following arguments. Polymers are stretched and rotated under the action of the local shear and tend on average to align with the streamlines, while the entropic forces acting to return the molecule to its undisturbed conformation lead to an extra tension in the direction of the flow.  Some well-known examples are the Weissenberg effect and die swell in fluid extrusion. In addition to large-scale collective effects, the presence of normal stress differences in flow can be important on smaller scales as well: cells and other soft biological matter may experience extra polymeric stresses that lead to deformation or possibly rupture \citep{Morozov2015}. Similar to cell migrations in blood vessels, researchers \citep{ho1976migration,halow1970experimental,d2010viscoelasticity} show that, due to the imbalanced normal stresses, in a simple shear flow particles close to the center plane of the setup tend to move toward the nearest wall. Many microorganisms swim through fluids that display non-Newtonian characteristics. For example, as spermatozoa make their journey through the female reproductive tract they encounter several complex fluids including glycoprotein-based cervical mucus in the cervix \citep{Katz1978}, mucosal epithelium inside the fallopian tubes, and an actin-based viscoelastic gel outside the ovum \citep{Dunn1976,Suarez2006}. These complex fluids often have dramatic effects on the locomotion of microorganisms. The presence of time-dependent stresses, normal stress differences, and shear-dependent material functions in complex fluids are able to fundamentally alter the physics of locomotion \citep{Purcel1977,Lauga2009}.

Propulsive forces can also result from the secondary flows induced by non-Newtonian normal stress differences; a theoretical investigation that further exemplifies these complexities is that by \citet{normand08}. They considered a biologically inspired geometric example of a semi-infinite flapper performing reciprocal sinusoidal motion in a viscoelastic Oldroyd-B fluid in the absence of inertia. They showed explicitly that the reciprocal motion generates a net force on the flapper occurring at second order in the flapping amplitude, and disappearing in the Newtonian limit as dictated by the scallop theorem, but there was no time-average flow accompanying the net force generation. Also, \citet{pak10} reported on the discovery of a net fluid flow produced by the reciprocal flapping motion in an Oldroyd-B fluid. The net flow transport was seen to occur at fourth order in the flapping amplitude, and was due to normal-stress differences. The dependence of the pumping performance on the actuation and material parameters was characterized analytically, and the optimal pumping rate was determined numerically. Through this example, they therefore demonstrated explicitly the breakdown of the scallop theorem in complex fluids in the context of fluid pumping, and suggested the possibility of exploiting intrinsic viscoelastic properties of the medium for fluid transport on small scales.

The investigation by \citet{Pak2012} is very relevant for the present paper. They reported that a two-sphere rotating dimer (snowman geometry) was capable of self-propelling in a complex fluid if the two spheres were of different sizes. The motion results from the asymmetry and the presence of normal stress differences under rotational actuation. Physically, the direction in which such an object moves can be understood by means of the hoop stresses generated along curved streamlines. A secondary, purely elastic flow is created by each rotating sphere, contracting in along the equator of each sphere and flowing out of the poles. Because the spheres are unequal in size, hydrodynamic interactions due to this secondary flow are unbalanced leading to propulsion in the direction of the smallest sphere. \citet{Puente-Velazquez2019} verified these findings experimentally using a magnetic snowman immersed in a Boger-type fluid. Recently \citet{Binagia2021} studied a mathematical model of two linked spheres rotating in opposite directions, which is a force and torque free swimmer. For this configuration the swimming direction was found to be toward the larger sphere instead of the smaller one, which is opposite to what was previously found \citep{Pak2012,Puente-Velazquez2019}. In addition, the asymmetry between the head and tail of a helical swimmer was reported to be responsible for the swimming speed enhancement of helical swimmers in viscoelastic fluids \citep{Angeles2021}.

Other studies have also shown that a wall can break the symmetry of flow leading to the propulsion of a dimer with equal spheres \citep{Keim2012} and a three-sphere microswimmer \citep{Daddi2018}. Other related investigations include the effect of the hydrodynamic interactions between two neighboring microswimmers near a wall \citep{Li2014}, given that boundaries have been shown to induce order in collective flows of bacterial suspensions \citep{Woodhouse2012,Wioland2013,Wioland2016}, leading to potential applications in autonomous microfluidic systems \citep{Woodhouse2017}. 

\begin{figure}
\centering
\includegraphics[height=0.4\textwidth]{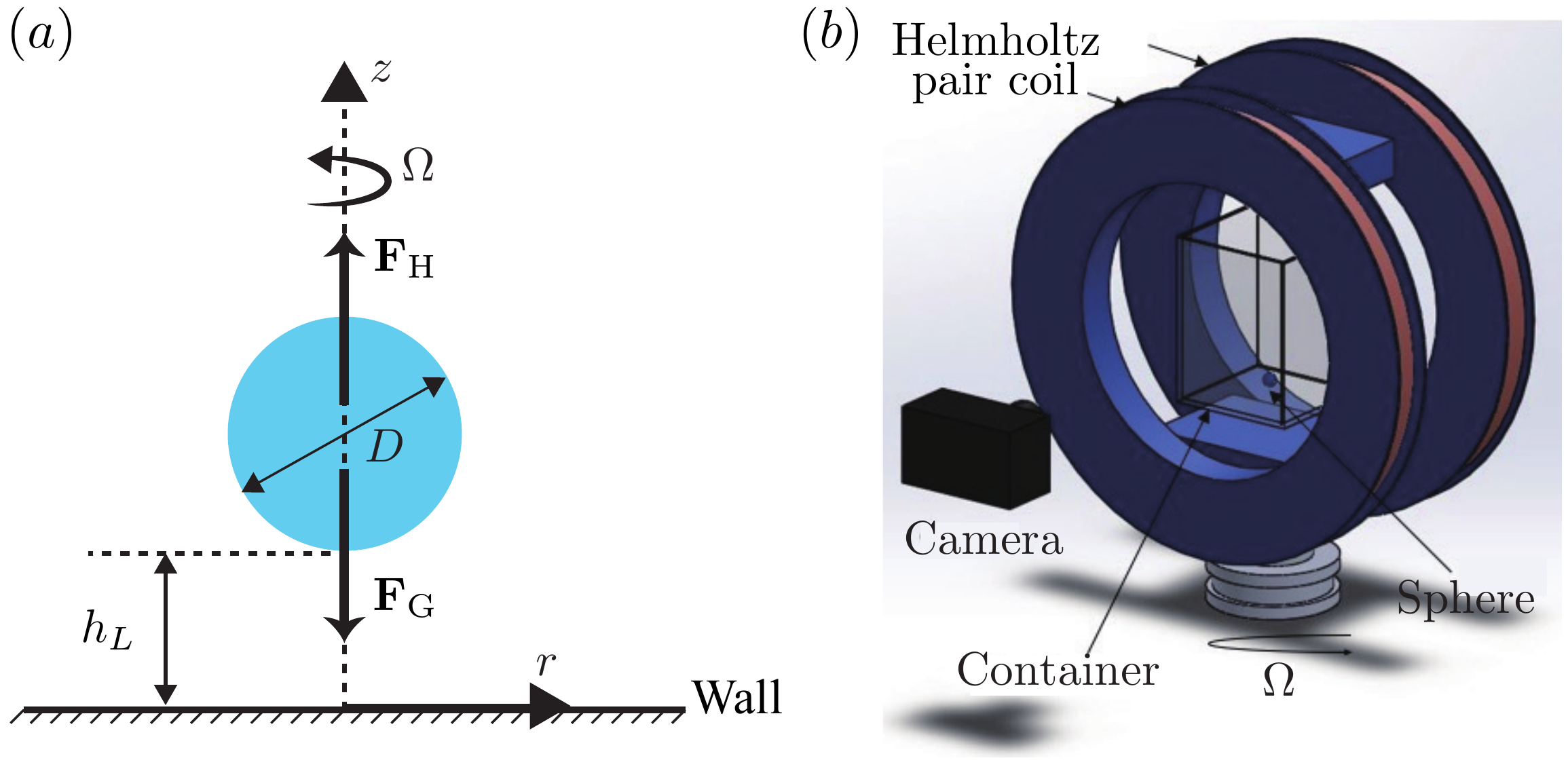} \
\caption{ (a) Schematic of a sphere of {diameter} $D$ {rotating} above a plane wall at a constant rotational rate
$\Omega $ about the $z$-axis. When the levitating hydrodynamic force $\mathbf{F}_\text{H}$ on the  sphere balances its own gravitational force $\mathbf{F}_\text{G}$, the  bottom of the sphere stays at a levitation height {$h=h_L$} above the wall. (b) {The experimental setup consists of a spherical particle inserted with permanent magnets placed inside a container of test fluid under a Helmholtz coil pair.}}
\label{fig:setup}
\end{figure}

In this work we introduce a novel phenomenon that arises from the effect of viscoelasticity via symmetry breaking. We experimentally observed that when a dense sphere is rotated near a wall being immersed in a viscoelastic fluid, it levitates to a fixed distance from the wall. We refer to this phenomena as `viscoelastic levitation'. The arrangement considered here is shown schematically in Fig.~\ref{fig:setup}a. Spheres of various sizes and densities were tested in a Boger-type fluid \citep{boger1977,James:2009} in experiments. We also develop a theoretical model that captures the dependence of the levitation height on the experimental parameters. A dimensionless group is identified to collapse the levitation results from  experiments.

\section{Experimental setup and test fluids}

\subsection{Experimental setup}
All experiments in this paper were conducted using the magnetic setup developed by \citet{Godinez2012} shown in Fig.~\ref{fig:setup}b. The device is capable of producing a magnetic field of 6 mT of uniform strength; the field is mechanically rotated. The spheres were placed inside a rectangular tank (160~mm $\times$ 100~mm $\times$ 100~mm) that fits into the region of uniform magnetic field inside the coils of approximately $100 \times 100 \times 100$~mm$^3$ in size where the test fluids were contained. The spheres were made out of plastic, inside which several permanent magnets were inserted (Magcraft, models NSN0658). For all the cases, the angular frequency of the rotating coils was below the step-out frequency~\citep{Godinez2012}; in other words, the sphere rotated at the same rate as the external magnetic field. 

Six spheres were tested. Table \ref{table:spheres} shows the properties of all spheres.  Two spheres (D2 and D3) had approximately the same diameter but different densities; and three spheres (D1, D2 and D4) had approximately the same density but different diameters. Two spheres (D5 and D6) had small densities but larger diameters. The sphere was initially placed at the bottom of the tank at rest and then driven by the external magnetic field to rotate with the rotating velocity vector normal to the horizontal plane wall (Fig.~\ref{fig:setup}). A camera was used to record the motion of the sphere rotating in the fluids and the recorded videos were used in the data analysis to track the vertical position of the sphere. 

\begin{table}
\begin{center}
\def~{\hphantom{0}}
    \begin{tabular}{cccc}
Sphere & Diameter, mm & Mass, mg & Density, kg/m\textsuperscript{3} \\ [3pt]
D1 (\protect\markerDone) & 7.99 & 585 & 2190.37 \\
D2 (\protect\markerDtwo) & 8.72 & 760 & 2189.10 \\
D3 (\protect\markerDthree) & 8.81 & 819 & 2287.48 \\ 
D4 (\protect\markerDfour) & 9.57 & 1005 & 2189.94 \\
D5 (\protect\markerDfive) & 13.0 & 1870 & 1626\\
D6 (\protect\markerDsix) & 16.0 & 3170 & 1478\\
\end{tabular}
\caption{Physical properties of the spheres used in this investigation.}
\label{table:spheres}
\end{center}
\end{table}

\subsection{Test fluids}

\begin{figure}
\centering
\includegraphics[height=0.9\textwidth]{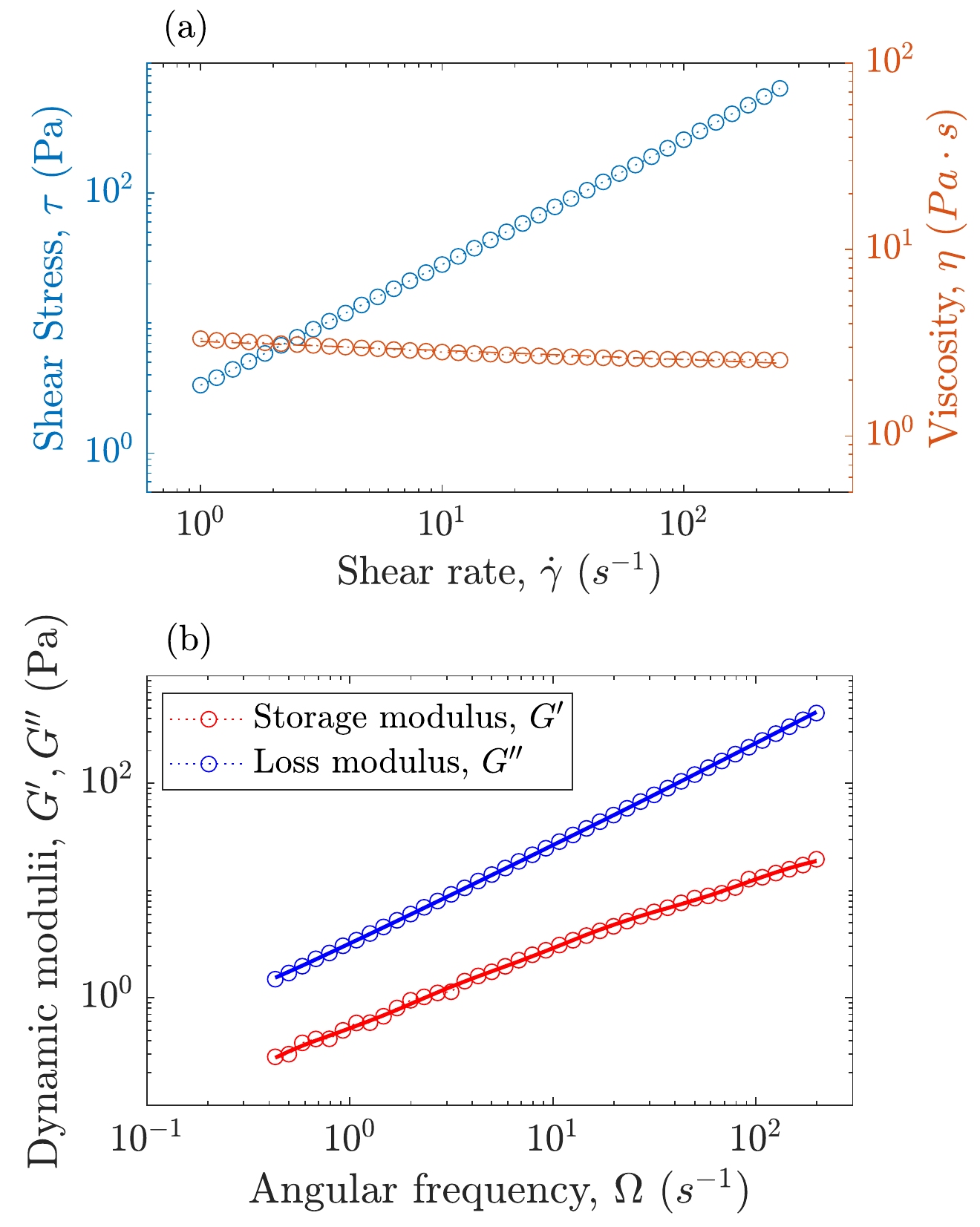} 
\
\caption{Rheology of the BF-II fluid: (a) shear stress vs shear rate for the viscosity measurement; (b) oscillation test for the relaxation time measurement, storage modulus (red circls) and loss modulus (blue circles) vs oscillating frequency $\Omega$. The solid lines show the fit to the data using the generalised Maxwell model Eq.~\ref{Eq:G'G''}.}
\label{fig:rheology}
\end{figure}

Two types of fluids were fabricated, tested, and used: one is Newtonian reference fluid (NF) and the other is Boger-type fluid (BF) (nearly constant shear viscosity but with viscoelastic properties). Table~\ref{table:rheology} summarizes the rheological properties of both fluids. To test the effect of changing the viscoelastic relaxation time, two different Boger fluids were prepared (BF-I and BF-II).

The Boger-type fluids were prepared by dissolving polyacrylamide (PAA, molecular weight 5 × $10^6$ g/mol) in nonionic water with slow mixing for 24 hours. Afterwards, the polymeric solution was added to a corn syrup solution with slow mixing over four days. 
The recipes (mass percentage of glucose, water and PAA) are (84.96$\%$, 15$\%$, 0.04$\%$) and (87.95$\%$, 12$\%$, 0.05$\%$), respectively. After the mixing the solution was left untouched for two weeks to remove the residual bubbles in the fluid before testing. The Newtonian fluid was made by mixing nonionic water with glucose and adjusting the percentage of water until the fluid showed a similar viscosity to the Boger fluid. All the fluids were stored and used in closed containers to avoid free surface crystallization. The rheological properties of the fluids were measured using a shear-rate controlled rheometer (Anton Paar, and ARES-G2, TA Instruments) with a cone-plate geometry. Both steady shear and oscillatory tests were conducted. Note that different batches of corn syrup were used to prepare BF-I ad BF-II. In both cases, the fluids had nearly constant viscosity and strong viscoelastic behaviour but their rheological characterization was different.

The details of the rheological characterization of the BF-I fluid can be found in \citet{Castillo2019}, but its salient features are summarized here.  The steady shear behaviour of this fluid was found to agree very well with the the Oldroyd-B model  \citep{oldroyd1950}. 
The measured first normal stress difference, $N_1$, agreed very closely to 
\begin{equation}\label{eqn:OlroydB}
   N_1=2\eta_0 (1-\zeta) \lambda {\dot\gamma}^2
 \end{equation}
where $\eta_0=\eta_p+\eta_s$ is the total viscosity (where $\eta_p$ and $\eta_s$ are the polymer and solvent viscosities, respectively) , $\zeta$ is s the ratio of solvent to total viscosities and $\lambda$ is the relaxation time. For the composition of the BF-I fluid we found that  $\zeta = 0.225$ and  $\eta_0=0.844$ Pa s, $\beta=0.225$. By fitting Eqn.(\ref{eqn:OlroydB}) to the rheological data, we obtain the relaxation time of the Boger fluid, $\lambda=0.51$ s.

The steady and oscillatory shear tests of the BF-II fluid are shown in Fig.~\ref{fig:rheology}. The fluid showed a nearly constant viscosity for the entire range of shear rates. The viscosity and the shear stress of the Boger fluid were fit to a power law model, leading to a power law index $n = 0.98$. Therefore, we consider the viscosity of the Boger fluid is effectively constant. The fist normal stress difference (not shown) was not quadratic with shear rate.

To find the relaxation time for the BF-II fluid, we used the oscillatority tests. Since there is no crossover of the storage modulus, G$'(\omega$), and loss modulus, G$''(\omega$), for this fluid, as shown in Fig. \ref{fig:rheology}b, the generalized Maxwell model was used to fit the experimental values of the G$'(\omega$) and G$''(\omega$) following \citet{baumgaertel1989}, \citet{liu2011force}, and \citet{espinosa2013fluid}.
The storage modulus and loss modulus are given by
\begin{equation}
  G'(\omega) = \sum_{i=1}^{N} \frac{g_i\lambda_i^2\omega^2}{1+\lambda_i^2\omega^2} \hspace{1.25cm}  \text{and} \hspace{1.25cm} G''(\omega) = \omega\eta + \sum_{i=1}^{N} \frac{g_i\lambda_i\omega}{1+\lambda_i^2\omega^2}
    \label{Eq:G'G''},
\end{equation} 
where $\omega$ is the oscillation frequency, $\eta$ is the viscosity of the Newtonian solvent and $g_i$ are the corresponding fitting parameters for relaxation time $\tau_i$. The corresponding relaxation time is determined by fitting the experimental data using Eq.~\ref{Eq:G'G''} with $N = 4$.

\begin{table}
 \begin{center}
\def~{\hphantom{0}}
  \begin{tabular}{cccccc}
Fluid & \(\rho\), kg/m\textsuperscript{3} &  $\eta_0$, Pa s &  Power law index, n & \( \lambda \), s  \\ [3pt]
Newtonian fluid (NF) & 1510 & 0.840 & 1.00 & 0.00 &  \\
Boger fluid I (BF-I) & 1508 & 0.844 & 0.96 & 0.51 \\
Boger fluid II (BF-II) & 1347 & 2.9 & 0.98 & 0.34\\
\end{tabular}
\caption{Physical properties of the fluids used in this investigation.}
\label{table:rheology}
\end{center}
\end{table}

\section{Results and discussion}

\subsection{Experimental results}
\begin{figure}
\centering
\includegraphics[width=0.9\textwidth]{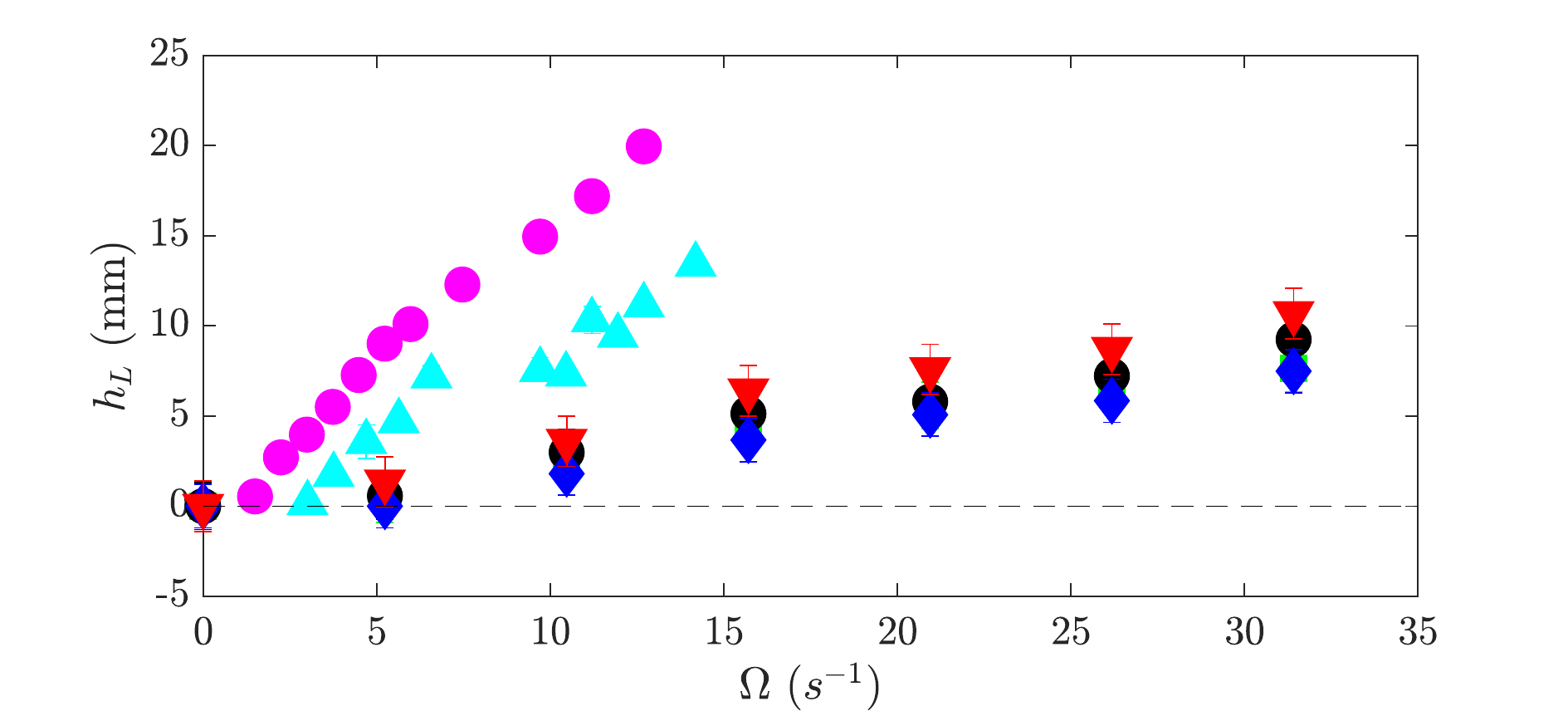}
\caption{Levitation height, $h_\text{L}$ (mm), as a function of the rotational speed, $\Omega$ (s\textsuperscript{-1}) for the Boger fluids (BF-I and BF-II). The symbols for each experiment correspond to those in Table \ref{table:spheres}. The dashed line shows the measurements for the Newtonian case (no levitation observed).}
\label{fig:exp}
\end{figure}

Figure~\ref{fig:exp} shows the experimental results of the levitation {height, $h_\text{L}$}, as a function of rotational speed, $\Omega$, for all the spheres tested in the Boger fluids.  In the case of Newtonian fluids (data not shown), the levitation distance is zero for all spheres and rotational speeds. This is expected since there is no shear-induced normal stress generated for Newtonian fluids. 
When  experiments were conducted with the spheres immersed in the Boger fluids (BF-I and BF-II), a significant levitation height, {$h_\text{L}$}, was observed, with the errorbars showing the variations of the levitation motion in the equilibrium state (gravity force balanced by levitating force). Videos of associated experiments can be found in the supplemental materials. In general, the levitation height increases with the rotating speed $\Omega$, indicating that there is a significant viscoelastic reaction from the fluid as a result of the rotation-induced shear in the gap between the sphere and the wall. Clearly, the levitation is solely a result of the viscoelastic nature of the fluid. The Reynolds number based on the rotating speed ranges from 0.5 to 3, for which inertial effects are small.

In particular, from the data shown in Fig. \ref{fig:exp}, we can see that for spheres of the same diameter (D2,\protect\markerDtwo, and D3,\protect\markerDthree), the levitation height is larger for the sphere of a smaller density (D2,\protect\markerDtwo ) at the same rotation rate; for spheres of the same density (D1, \protect\markerDone, and D4,\protect\markerDfour ), the levitation height is larger for the sphere of a larger diameter (D4,\protect\markerDfour) considering the same rotational speed.
To understand the levitation height dependence on the experimental parameters (density and diameter), we compose a theoretical model that can be compared with the experimental results. The model, however, is valid only for small values of De.

\subsection{Theoretical model}

We consider a sphere of diameter $D$ rotating at a constant velocity $\Omega$ 
near an infinitely large wall (see Fig.~\ref{fig:setup}a). The rotational axis 
is along the wall-normal direction ($z$) and the bottom of sphere is above the 
wall by $h$. Hence, the configuration is axisymmetric and can be described by 
the $rz$ cylindrical coordinates. 
The density of sphere is assumed to be larger than that of the 
carrier fluid, hence their density difference $\Delta \rho > 0$. We use the the Oldroyd-B constitutive model to capture the viscoelasticity of the fluid, which was shown to agree well with the rheological behavior of fluid BF-I. Although the Oldroyd-B model does not predict the second normal stress difference, the magnitude of the second normal stress difference is typically much smaller compared with that of the first normal stress difference, making the Oldroyd-B model a reasonable approximation of a Boger fluid. The governing 
equations of the fluid are 
\begin{align}
\boldsymbol{\nabla} \cdot \bu & = 0, \nonumber \\
\boldsymbol{\nabla} \cdot \bsig & = \mathbf{0},
\end{align}
where $\bsig = -p \bI + \eta_s \strain + \taup$, 
$p$ and $\bu$ denote the pressure and velocity respectively, $\strain = \boldsymbol{\nabla} \bu + \left(\boldsymbol{\nabla} 
\bu\right)^T$ denotes the rate of strain tensor. The relative viscosity $\zeta = \eta_s/\eta_0 < 1$ is defined as the ratio between $\eta_s$ and the total viscosity 
$\eta_0$. 
The polymeric stress $\taup$ is governed by the upper-convected Maxwell equation:
\begin{align}
\lambda \overset{\triangledown}{\taup} + \taup = \eta_p \strain ,
\end{align}
where the upper-convected derivative $\overset{\triangledown}{\bA}$ on a tensor $\bA$ is defined as $\overset{\triangledown}{\bA} = \partial \bA/\partial t + \bu \cdot 
\boldsymbol{\nabla} \bA - \left(\boldsymbol{\nabla}\bu\right)^T \cdot \bA - \bA \cdot \boldsymbol{\nabla} \bu$. Here $\lambda$ denotes the relaxation time of the polymeric fluid, and the polymeric viscosity $\eta_p = 
(1-\zeta) \eta_0$. 

Due to axisymmetry, the levitating force due to the viscoelastic stress $\mathbf{F}_\text{H}$ is along the $z$ 
direction, which should balance the gravity-induced force $\mathbf{F}_\text{G} = 
-\frac{\pi}{6}\Delta \rho g D^3\bez$, where $g$ denotes the gravitational 
acceleration. {For a given polymeric fluid and a given rotational speed, we seek a levitation height $h_\text{L}$ such that
$\mathbf{F}_\text{H}(\Omega,h=h_\text{L}) = - \mathbf{F}_\text{G}$, when the rotating
sphere suspends above the wall by a finite distance, with $h_\text{L}>0$.}

\subsubsection{Nondimensionalization}
We scale lengths by $D$, time by $1/\Omega$, velocities by $\Omega D$, and stresses by $\eta_0 \Omega$, with the nondimensional variables denoted with tildes.
The nondimensional governing equations are therefore given by 
\begin{align}\label{eq:non1}
 \nnabla  \cdot \tilde{\mathbf{u}} & = 0, \nonumber \\
 -\nnabla \tilde{p} + \zeta \nnabla^2 \tilde{\mathbf{u}} + \nnabla \cdot \tilde{\boldsymbol{\tau}}_p & = \mathbf{0}, \nonumber \\
 \De \tilde{\boldsymbol{\tau}}_p + \overset{\triangledown}{\tilde{\boldsymbol{\tau}}}_p & = (1-\zeta)\tilde{\strain},
\end{align}
where $\De = \lambda \Omega$ is the Deborah number indicating the 
nondimensional relaxation time of the viscoelastic fluid. We hence seek
a nondimensional levitation height ($\tilde{h}=\tilde{h}_\text{L}$) such that 
\begin{align}
\tilde{F}_\text{H}\lp \De, \tilde{h}_\text{L} \rp= \text{G}, \label{GB}
\end{align}
where 
\begin{align}
\text{G}  = \frac{\pi g D \Delta \rho}{6 \eta_0 \Omega}
\end{align}
is the dimensionless gravitational force.

\subsubsection{Small Deborah number analysis: a reciprocal theorem approach}
We first consider the small-De limit of Eqs.~\ref{eq:non1} and adopt the second-order fluid model to describe the first departure from Newtonian behavior. 
In a retarded motion expansion, the nondimensional shear stress tensor of a second-order fluid reads
\begin{align}\label{eq:non_second}
\tilde{\boldsymbol{\tau}} = \tilde{\mathbf{E}} - \De_0 \lp  \overset{\triangledown}{\tilde{\mathbf{E}}} - 
\frac{2\Psi_2}{\Psi_1} \tilde{\mathbf{E}} \cdot \tilde{\mathbf{E}}  \rp,
\end{align}
where $\Psi_1$ and $\Psi_2$ are the first and second normal stress 
coefficients, respectively. Here, $\De_0 = \Psi_1 \Omega/\eta$ defines the Deborah number of the second-order fluid, and it relates to $\De$ by $\De_0 = (1-\zeta)\De$. For comparison with the Oldroyd-B model, where the second normal stress difference is zero, we set $\Psi_2=0$ to recover the Oldroyd-B model in the small-$\De$ limit.

\begin{figure}
\centering
	\includegraphics[width=1\columnwidth]
	{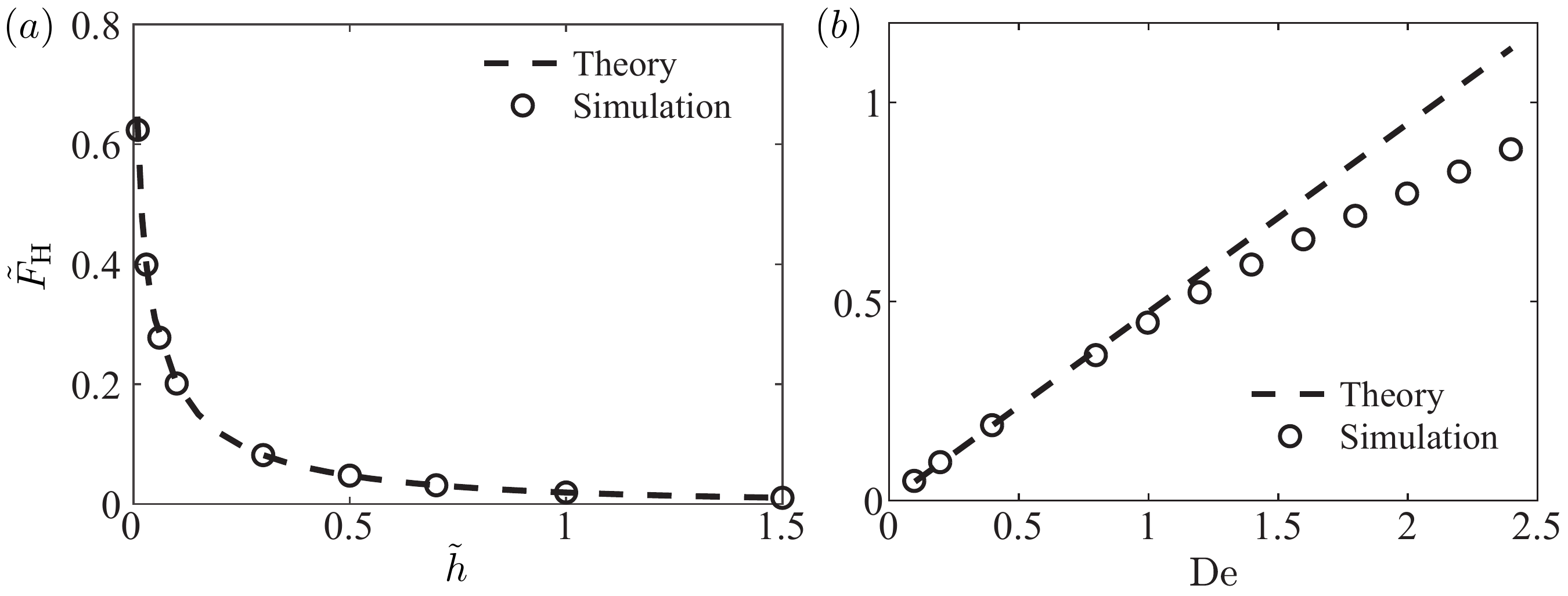}
\caption{Nondimensional hydrodynamic force $\tilde{F}_\text{H}$ on a rotating sphere as a function of (a) its nondimensional height $\tilde{h}$ from the wall when $\De=0.1$, and (b) Deborah number $\De$ at a fixed height $\tilde{h}=1$. In both cases, $\zeta = 0.225$. Lines and circles denote the theoretical and numerical 
results, respectively.}
\label{fig:force_valid}
\end{figure}

We first  asymptotically calculate the hydrodynamic force $\tilde{\mathbf{F}}_\text{H} =  \tilde{F}_\text{H} \bez$ on a rotating sphere suspended at a given height, $\tilde{h}$. 
We expand the variables in powers of $\De_0$ as
\begin{align}
\{\tilde{\boldsymbol{\sigma}}, \tilde{\mathbf{u}}, \tilde{\mathbf{E}}, \tilde{\mathbf{F}}_\text{H} \} =\{\tilde{\boldsymbol{\sigma}}_0, \tilde{\mathbf{u}}_0, \tilde{\mathbf{E}}_0, \tilde{\mathbf{F}}_0 \}+\De_0\{\tilde{\boldsymbol{\sigma}}_1, \tilde{\mathbf{u}}_1, \tilde{\mathbf{E}}_1, \tilde{\mathbf{F}}_1 \}+ O \lp \Des^2 \rp.
\end{align}
The zeroth-order solution $\left\{ \tilde{\mathbf{u}}_0, \tilde{\boldsymbol{\sigma}}_0 = -\tilde{p}_0 \bI + \tilde{\mathbf{E}}_0 
\right\}$ is a known Newtonian (Stokes flow) solution for a rotating sphere above a wall~\citep{jeffery1915steady}, where the zeroth-order hydrodynamic force on the sphere $\tilde{\mathbf{F}}_0 = 
\mathbf{0}$. Levitation of a rotating sphere near wall is therefore impossible in a Newtonian fluid.

Next we calculate the first-order non-Newtonian correction $\{ \tilde{\mathbf{u}}_1, \tilde{\boldsymbol{\sigma}}_1 = -\tilde{p}_1 
\bI + \tilde{\mathbf{E}}_1 - \overset{\triangledown}{\tilde{\mathbf{E}}}_0 \}$ via a 
reciprocal theorem approach \citep{Lauga2014,elfring_2017,masoud_stone_2019}. By considering an auxiliary problem in Stokes flow $(\tilde{\mathbf{u}}', \tilde{\boldsymbol{\sigma}}')$ where a sphere translates perpendicularly to a wall, which has an exact solution given by \citet{brenner1961}, the reciprocal theorem leads to  
\begin{align}\label{eq:recip1}
 \int_V \lp \tilde{\boldsymbol{\nabla}} \tilde{\mathbf{u}}': \tilde{\boldsymbol{\sigma}}_1   - \tilde{\boldsymbol{\nabla}} \tilde{\mathbf{u}}_1 : \tilde{\boldsymbol{\sigma}}' \rp \d V = \int_V 
\tilde{\boldsymbol{\nabla}} \cdot \lp \tilde{\mathbf{u}}' \cdot \tilde{\boldsymbol{\sigma}}_1 -  \tilde{\mathbf{u}}_1 \cdot \tilde{\boldsymbol{\sigma}}' \rp \d V.
\end{align}
Upon the substitution of the first-order constitutive equation $\tilde{\boldsymbol{\sigma}}_1 = -\tilde{p}_1 
\bI + \tilde{\mathbf{E}}_1 - \overset{\triangledown}{\tilde{\mathbf{E}}}_0$ and the use of the divergence theorem, we obtain
\begin{align}\label{eq:recip2}
 \int_V \overset{\triangledown}{\tilde{\mathbf{E}}}_0 : \tilde{\boldsymbol{\nabla}} \tilde{\mathbf{u}}' \d V = \int_{S}\bn 
\cdot \lp \tilde{\mathbf{u}}' \cdot \tilde{\boldsymbol{\sigma}}_1 - \tilde{\mathbf{u}}_1 \cdot \tilde{\boldsymbol{\sigma}}' \rp \d S,
\end{align}
where the surface integral on the stationary wall vanishes due to the no-slip and no-penetration boundary conditions, and $S$ and $\bn$ denote the surface of the sphere and its outward 
normal, respectively. In Eq.~\ref{eq:recip2}, the first order velocity on the surface $S$ vanishes because the rotational velocity has been accounted by in the zeroth-order solution and a fixed distance from the wall is considered here. Furthermore, by considering a sphere translating at a unit speed  $\tilde{\mathbf{u}}'  = \bez$ in the auxiliary problem, Eq.~\ref{eq:recip2} is simplified to
\begin{align}
 \tilde{F}_1 =  -\int_V  
\overset{\triangledown}{\tilde{\mathbf{E}}}_0 
: \nnabla \tilde{\mathbf{u}}'  \mathrm{d}V, \label{eqn:asym}
\end{align}
where $\tilde{F}_1=\bez \cdot \tilde{\mathbf{F}}_1 = \bez \cdot \int_{S} \lp -\bn \cdot \tilde{\boldsymbol{\sigma}}_1 \rp \d S$ represents the first-order levitating force. In other words, the leading order levitating force therefore reads
\begin{align}
 \tilde{F}_H = \De_0 \tilde{F}_1 = -\De \lp 1- \zeta \rp \int_V  
\overset{\triangledown}{\tilde{\mathbf{E}}}_0 
: \nnabla \tilde{\mathbf{u}}'  \mathrm{d}V. \label{eqn:Fhydro}
\end{align} 
The above analysis, valid in the small-De regime, provides the theoretical foundation for the levitation of a rotating sphere in a viscoelastic fluid.

For illustration, the levitating force $\tilde{F}_\text{H}$ is calculated as a function of distance from the wall $\tilde{h}$ at fixed $\De=0.1$ and $\zeta=0.225$ (Fig.~\ref{fig:force_valid}a, dashed line). The levitating force decays as the rotating sphere is further away from the wall. For verifications, $\tilde{F}_\text{H}$ is also computed numerically using a commercial 
finite-element solver COMSOL based on our legacy implementation~\citep{laipof1,Pak2012, nadal2014rotational, datt2015squirming}. The numerical results (represented by circles in Fig.~\ref{fig:force_valid}a) display excellent agreements with the asymptotic solution for $\De=0.1$. We remark that both Newtonian solutions \citep{jeffery1915steady, brenner1961} employed in Eq.~\ref{eqn:Fhydro} are series solutions. Although the solutions are valid for all distance above the wall, as the sphere gets closer to the wall, an increasingly higher number of terms are required in the series for accurate solutions. We therefore limit our consideration of the distance to $\tilde{h}>0.01$ in this work.

In Fig.~\ref{fig:force_valid}b, we test the effect of higher Deborah number numerically (circles) and compare with asymptotic solution (solid line). At a fixed distance from the wall, the levitating force increases with $\De$. The asymptotic solution displays excellent agreements with the numerical results up to $\De \approx 1$, beyond which the asymptotic solution overestimates the levitating force, which is reasonable considering the small De assumption in the asymptotic analysis. We note that currently we have no access to numerical results at even higher $\De$ due to the limitations by the high Weissenberg number problem \citep{keunings1986,owens2002}.

\subsection{Determination of the levitation height}
From the levitating force on the rotating sphere as a function of its distance from the wall, we can determine the levitation height of the sphere at which the levitating force  balances the gravitational force. Substituting the leading-order viscoelastic force in the small Deborah number limit given by Eq.~\ref{eqn:Fhydro}, $\tilde{F}_H (\De, \tilde{h}) \sim \text{De}(1-\zeta) \tilde{F}_1 (\tilde{h})$, into the force balance  (Eq.~\ref{GB}), we have
\begin{align}
\De (1-\zeta) \tilde{F}_1 (\tilde{h}=\tilde{h}_\text{L}) &= \text{G},
\end{align}
which, upon bringing the relevant dimensionless groups together, yields
\begin{align}
\tilde{F}_1 (\tilde{h}=\tilde{h}_\text{L}) &= \frac{\text{G}}{\De (1-\zeta)} \cdot \label{eqn:FinalBalance}
\end{align}
Therefore, the solution for the nondimensional levitation height, $\tilde{h}_L$, in the above force balance should only depend on the dimensionless group, $\text{De}(1-\zeta)/\text{G}$, in the regime of small $\text{De}$. For a given value of $\text{De}(1-\zeta)/\text{G}$, we obtain the solution $\tilde{h}=\tilde{h}_L$  by evaluating numerically the levitating force using Eq.~\ref{eqn:asym} such that Eq.~\ref{eqn:FinalBalance} is satisfied. We remark that the De numbers in the experiments are typically large, so the asymptotic theory is not expected to quantitatively capture the experimental measurements. Instead, the asymptotic analysis here serves only to predict the plausibility of viscoelastic levitation and suggest the relevant dimensionless group $\text{De}(1-\zeta)/\text{G}$ for collapsing the data in the asymptotic regime of small $\text{De}(1-\zeta)/\text{G}$.

\begin{figure}
\centering
	\includegraphics[width=1\columnwidth]
	{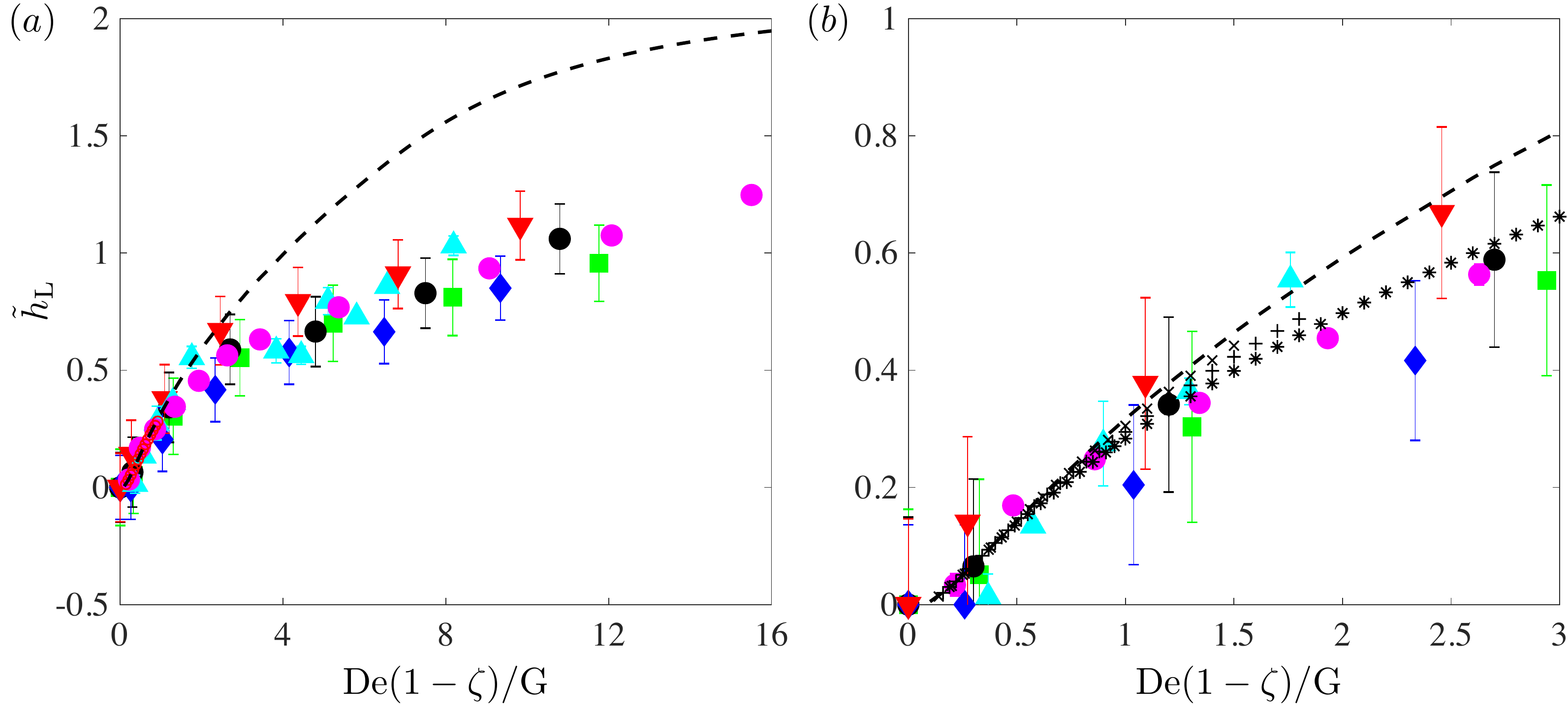}
\caption{(a) Dimensionless levitation height $\tilde{h}_\text{L} = h_\text{L}/D$ of the rotating sphere as a function of the dimensionless group, $\De(1-\zeta)/\text{G}$. The dashed line represents predictions by the asymptotic theory in the small Deborah number limit, whereas the symbols correspond to experimental data previously presented. (b) A magnified view of results in (a) for small values of $\text{De}(1-\zeta)/G$, with the addition of results from numerical simulations for $\text{De}=1$ ($\times$), $\text{De}=1.5$ ($+$), and $\text{De}=2$ ($*$); in all simulations, $\zeta=0.225$.}
\label{fig:illustrate}
\end{figure}

Figure \ref{fig:illustrate} shows the nondimensional levitation height ($\tilde{h}_\text{L}=h_\text{L}/D$) from both asymptotic theory predictions (dashed lines) and experimental measurements (symbols) as a function of the dimensionless group, $\text{De}(1-\zeta)/\text{G}$. 
It is to be noted that there is a good agreement between the asymptotic and experimental results when $\text{De}(1-\zeta)/\text{G}$ is small. At larger $\text{De}(1-\zeta)/\text{G}$, the asymptotic theory over estimates the levitation height, which is due to the small $\De$ assumption in the asymptotic theory. This over-prediction at large $\De$ can also be seen in the force comparison between the asymptotic theory and the numerical simulations in Fig.~\ref{fig:force_valid}b. In Fig.~\ref{fig:illustrate}b, we show a magnified view of results for small values of $\text{De}(1-\zeta)/\text{G}$. In addition, we superimpose results from numerical simulations for $\text{De}=1$ ($\times$), $\text{De}=1.5$ ($+$), and $\text{De}=2$ ($*$) for comparison. We can see that the numerical results agree well with the asymptotic theory when $\text{De}(1-\zeta)/\text{G}$ is small; in this regime, the data collapse well, confirming that the levitation height depends only on the dimensionless group of $\text{De}(1-\zeta)/\text{G}$.

\section{Conclusions}
In this study, we conducted experiments and {theoretical} analysis on spheres of different sizes and densities immersed in two fluids: Newtonian and viscoelastic Boger fluids. 
With a constant rotating rate, the sphere levitates to a fixed distance from the bottom in the viscoelastic fluid, instead of no levitation in Newtonian fluids. The viscoelastic normal stress between the sphere and the bottom wall is responsible for this `viscoleastic levitation'. In the small-De asymptotic analysis, based on the balance between the {viscoelastic levitating force and the gravitational force on the sphere}, a dimensionless group was formulated in terms of the Deborah number $\De$, {the relative viscosity $\zeta$}, and the gravity number $\text{G}$. Using this dimensionless group, {experimental measurements of the levitation height display a good collapse onto a single curve.} The agreement between experiments and asymptotic results is {very good} when $\De$ is small, consistent with the small $\De$ assumption {in the asymptotic analysis.}

It can be argued that this configuration can be used as a rheometer. If the density and size of a sphere that rotates above a wall are known, a measurement of the levitation height can be used to infer the value of the Deborah number, from which the fluid relaxation time could be obtained. This method could be easily implemented considering the experimental device shown here, for small Deborah numbers. Other rotation directions and non-constant rotation speeds could also be considered to obtain other viscoelastic characteristics of the fluid. We plan to pursue these ideas in the future.

\begin{acknowledgments}

A. C. gratefully acknowledges financial support from Consejo Nacional de Ciencia y Tecnolog\'ia (M\'exico) through the scholarship no. 416397. L. Z. thanks the start-up grant provided by the National University of Singapore (R-265-000-696-133) and the A*STAR AME YIRG grant (A2084c0175). O. S. P. acknowledges support by the National Science Foundation (Grant No.~CBET-1931292).
\end{acknowledgments}

Declaration of Interests. The authors report no conflict of interest.



\begin{thebibliography}{41}
\providecommand{\natexlab}[1]{#1}
\providecommand{\url}[1]{\texttt{#1}}
\expandafter\ifx\csname urlstyle\endcsname\relax
  \providecommand{\doi}[1]{doi: #1}\else
  \providecommand{\doi}{doi: \begingroup \urlstyle{rm}\Url}\fi

\bibitem[Larson(1999)]{Larson:1999}
R.~G. Larson.
\newblock \emph{The Structure and Rheology of Complex Fluids}.
\newblock Oxford University Press, Oxford, 1999.

\bibitem[Morozov and Spagnolie(2015)]{Morozov2015}
A.~Morozov and S.~E. Spagnolie.
\newblock Introduction to complex fluids.
\newblock In S.~E. Spagnolie, editor, \emph{Complex Fluids in Biological
  Systems. Experiment, Theory, and Computation}, chapter~1, pages 3--52.
  Springer, New York, NY, 2015.

\bibitem[d’Avino et~al.(2010)d’Avino, Maffettone, Greco, and
  Hulsen]{d2010viscoelasticity}
G~d’Avino, PL~Maffettone, F~Greco, and MA~Hulsen.
\newblock Viscoelasticity-induced migration of a rigid sphere in confined shear
  flow.
\newblock \emph{Journal of Non-Newtonian Fluid Mechanics}, 165\penalty0
  (9-10):\penalty0 466--474, 2010.

\bibitem[Halow and Wills(1970)]{halow1970experimental}
JS~Halow and GB~Wills.
\newblock Experimental observations of sphere migration in couette systems.
\newblock \emph{Industrial \& Engineering Chemistry Fundamentals}, 9\penalty0
  (4):\penalty0 603--607, 1970.

\bibitem[Ho and Leal(1976)]{ho1976migration}
BP~Ho and LG~Leal.
\newblock Migration of rigid spheres in a two-dimensional unidirectional shear
  flow of a second-order fluid.
\newblock \emph{Journal of Fluid Mechanics}, 76\penalty0 (4):\penalty0
  783--799, 1976.

\bibitem[Katz et~al.(1978)Katz, Mills, and Pritchett]{Katz1978}
D.~F. Katz, R.~N. Mills, and T.~R. Pritchett.
\newblock The movement of human spermatozoa in cervical mucus.
\newblock \emph{J. Reprod. Fertil.}, 53:\penalty0 259--265, 1978.

\bibitem[Dunn and Picologlou(1976)]{Dunn1976}
P.~F. Dunn and B.~F. Picologlou.
\newblock Viscoelastic properties of cumulus oophorus.
\newblock \emph{Biorheol.}, 13:\penalty0 379--384, 1976.

\bibitem[Suarez and Pacey(2006)]{Suarez2006}
S.~S. Suarez and A.~A. Pacey.
\newblock Sperm transport in the female reproductive tract.
\newblock \emph{Human Reprod. Update}, 1:\penalty0 23--37, 2006.

\bibitem[Lauga and Powers(2009)]{Lauga2009}
E.~Lauga and T.~R. Powers.
\newblock The hydrodynamics of swimming microorganisms.
\newblock \emph{Rep. Prog. Phys.}, 72:\penalty0 096601, 2009.

\bibitem[Purcell(1977)]{Purcel1977}
E.~M. Purcell.
\newblock Life at low reynolds number.
\newblock \emph{Am. J. Phys.}, 45:\penalty0 3--11, 1977.

\bibitem[Normand and Lauga(2008)]{normand08}
T.~Normand and E.~Lauga.
\newblock Flapping motion and force generation in a viscoelastic fluid.
\newblock \emph{Phys. Rev E}, 78:\penalty0 061907, 2008.

\bibitem[Pak et~al.(2010)Pak, Normand, and Lauga]{pak10}
O.~S. Pak, T.~Normand, and E.~Lauga.
\newblock Pumping by flapping in a viscoelastic fluid.
\newblock \emph{Phys. Rev. E}, 81:\penalty0 036312, Mar 2010.

\bibitem[Pak et~al.(2012)Pak, Zhu, Brandt, and Lauga]{Pak2012}
O.~S. Pak, L.~Zhu, L.~Brandt, and E.~Lauga.
\newblock Micropropulsion and microrheology in complex fluids via symmetry
  breaking.
\newblock \emph{Phys. Rev. Fluids}, 24:\penalty0 103102, 2012.

\bibitem[Puente-Vel\'azquez et~al.(2019)Puente-Vel\'azquez, God\'inez, Lauga,
  and Zenit]{Puente-Velazquez2019}
J.~A. Puente-Vel\'azquez, F.~A. God\'inez, E.~Lauga, and R.~Zenit.
\newblock Viscoelastic propulsion of a rotating dumbbell.
\newblock \emph{Microfluidics and Nanofluidics}, 23:\penalty0 108, 2019.

\bibitem[Binagia and Shaqfeh(2021)]{Binagia2021}
J.~P. Binagia and E.~S.~G. Shaqfeh.
\newblock Self-propulsion of a freely suspended swimmer by a swirling tail in a
  viscoelastic fluid.
\newblock \emph{Phys. Rev. Fluids}, 2021.

\bibitem[Angeles et~al.(2021)Angeles, God\'{\i}nez, Puente-Velazquez,
  Mendez-Rojano, Lauga, and Zenit]{Angeles2021}
Veronica Angeles, Francisco~A. God\'{\i}nez, Jhonny~A. Puente-Velazquez,
  Rodrigo Mendez-Rojano, Eric Lauga, and Roberto Zenit.
\newblock Front-back asymmetry controls the impact of viscoelasticity on
  helical swimming.
\newblock \emph{Phys. Rev. Fluids}, 6:\penalty0 043102, Apr 2021.
\newblock \doi{10.1103/PhysRevFluids.6.043102}.

\bibitem[Keim et~al.(2012)Keim, Garcia, and Arratia]{Keim2012}
N.~C. Keim, M.~Garcia, and P.~E. Arratia.
\newblock Fluid elasticity can enable propulsion at low reynolds number.
\newblock \emph{Phys. Fluids}, 24:\penalty0 081703, 2012.

\bibitem[Daddi-Moussa-Ider et~al.(2018)Daddi-Moussa-Ider, Lisicki, Hoell, and
  Lowen]{Daddi2018}
A.~Daddi-Moussa-Ider, M.~Lisicki, C.~Hoell, and H~Lowen.
\newblock Swimming trajectories of a three-sphere microswimmer near a wall.
\newblock \emph{J. Chem. Phys.}, 148:\penalty0 134904, 2018.

\bibitem[Li and Ardekani(2014)]{Li2014}
G.-J. Li and A.~M. Ardekani.
\newblock Hydrodynamic interaction of microswimmers near a wall.
\newblock \emph{Phys. Rev. E}, 90:\penalty0 013010, 2014.

\bibitem[Wioland et~al.(2016)Wioland, Woodhouse, Dunkel, and
  Goldstein]{Wioland2016}
H.~Wioland, F.~G. Woodhouse, J.~Dunkel, and R.~E. Goldstein.
\newblock Ferromagnetic and antiferromagnetic order in bacterial vortex
  lattices.
\newblock \emph{Nat. Phys.}, 12:\penalty0 341–--345, 2016.

\bibitem[Wioland et~al.(2013)Wioland, Woodhouse, Dunkel, Kessler, and
  Goldstein]{Wioland2013}
H.~Wioland, F.~G. Woodhouse, J.~Dunkel, J.~O. Kessler, and R.~E. Goldstein.
\newblock Confinement stabilizes a bacterial suspension into a spiral vortex.
\newblock \emph{Phys. Rev. Lett.}, 110:\penalty0 268102, 2013.

\bibitem[Woodhouse and Goldstein(2012)]{Woodhouse2012}
F.~G. Woodhouse and R.~E. Goldstein.
\newblock Spontaneous circulation of confined active suspensions.
\newblock \emph{Phys. Rev. Lett.}, 109:\penalty0 168105, 2012.

\bibitem[Woodhouse and Dunkel(2017)]{Woodhouse2017}
F.~G. Woodhouse and J.~Dunkel.
\newblock Active matter logic for autonomous microfluidics.
\newblock \emph{Nat. Commun.}, 8:\penalty0 15169, 2017.

\bibitem[Boger(1977)]{boger1977}
D.~V. Boger.
\newblock A highly elastic constant-viscosity fluid.
\newblock \emph{J. Non-Newton. Fluid Mech.}, 3:\penalty0 87--91, 1977.

\bibitem[James(2009)]{James:2009}
D.~James.
\newblock Boger fluids.
\newblock \emph{Ann. Rev. Fluid Mech.}, 41:\penalty0 129--142, 2009.

\bibitem[God\'inez et~al.(2012)God\'inez, Ch\'avez, and Zenit]{Godinez2012}
F.A. God\'inez, O.~Ch\'avez, and R.~Zenit.
\newblock Note: Design of a novel rotating magnetic field device.
\newblock \emph{Rev. Sci. Instrum.}, 83:\penalty0 066109, 2012.

\bibitem[Castillo et~al.(2019)Castillo, Murch, Einarsson, Mena, Shaqfeh, and
  Zenit]{Castillo2019}
Alfonso Castillo, William~L. Murch, Jonas Einarsson, Baltasar Mena, Eric S.~G.
  Shaqfeh, and Roberto Zenit.
\newblock Drag coefficient for a sedimenting and rotating sphere in a
  viscoelastic fluid.
\newblock \emph{Phys. Rev. Fluids}, 4:\penalty0 063302, Jun 2019.
\newblock \doi{10.1103/PhysRevFluids.4.063302}.
\newblock URL \url{https://link.aps.org/doi/10.1103/PhysRevFluids.4.063302}.

\bibitem[Oldroyd(1950)]{oldroyd1950}
J.~G. Oldroyd.
\newblock On the formulation of rheological equations of state.
\newblock \emph{Proc. Roy. Soc.}, 200:\penalty0 523--541, 1950.

\bibitem[Baumgaertel and Winter(1989)]{baumgaertel1989}
M.~Baumgaertel and H.~H. Winter.
\newblock Determination of discrete relaxation and retardation time spectra
  from dynamic mechanical data.
\newblock \emph{Rheol. Acta}, 28:\penalty0 511--519, 1989.

\bibitem[Liu et~al.(2011)Liu, Powers, and Breuer]{liu2011force}
Bin Liu, Thomas~R Powers, and Kenneth~S Breuer.
\newblock Force-free swimming of a model helical flagellum in viscoelastic
  fluids.
\newblock \emph{Proceedings of the National Academy of Sciences}, 108\penalty0
  (49):\penalty0 19516--19520, 2011.

\bibitem[Espinosa-Garcia et~al.(2013)Espinosa-Garcia, Lauga, and
  Zenit]{espinosa2013fluid}
Julian Espinosa-Garcia, Eric Lauga, and Roberto Zenit.
\newblock Fluid elasticity increases the locomotion of flexible swimmers.
\newblock \emph{Physics of Fluids}, 25\penalty0 (3):\penalty0 031701, 2013.

\bibitem[Jeffery(1915)]{jeffery1915steady}
G.~B. Jeffery.
\newblock On the steady rotation of a solid of revolution in a viscous fluid.
\newblock \emph{Proceedings of the London Mathematical Society}, 2\penalty0
  (1):\penalty0 327--338, 1915.

\bibitem[Elfring(2017)]{elfring_2017}
Gwynn~J. Elfring.
\newblock Force moments of an active particle in a complex fluid.
\newblock \emph{J. Fluid Mech.}, 829:\penalty0 R3, 2017.

\bibitem[Lauga(2014)]{Lauga2014}
Eric Lauga.
\newblock Locomotion in complex fluids: Integral theorems.
\newblock \emph{Phys. Fluids}, 26\penalty0 (8):\penalty0 081902, 2014.

\bibitem[Masoud and Stone(2019)]{masoud_stone_2019}
Hassan Masoud and Howard~A. Stone.
\newblock The reciprocal theorem in fluid dynamics and transport phenomena.
\newblock \emph{J. Fluid Mech.}, 879:\penalty0 P1, 2019.
\newblock \doi{10.1017/jfm.2019.553}.

\bibitem[Brenner(1961)]{brenner1961}
Howard Brenner.
\newblock The slow motion of a sphere through a viscous fluid towards a plane
  surface.
\newblock \emph{Chem. Eng. Sci.}, 16\penalty0 (3–4):\penalty0 242 -- 251,
  1961.

\bibitem[Datt et~al.(2015)Datt, Zhu, Elfring, and Pak]{datt2015squirming}
C.~Datt, L.~Zhu, G.~J. Elfring, and O.~S. Pak.
\newblock Squirming through shear-thinning fluids.
\newblock \emph{J. Fluid Mech.}, 784:\penalty0 R1, 2015.

\bibitem[Nadal et~al.(2014)Nadal, Pak, Zhu, Brandt, and
  Lauga]{nadal2014rotational}
F.~Nadal, O.~S. Pak, L.~Zhu, L.~Brandt, and E.~Lauga.
\newblock Rotational propulsion enabled by inertia.
\newblock \emph{Eur. Phys, J. E.}, 37\penalty0 (7):\penalty0 60, 2014.

\bibitem[Zhu et~al.(2012)Zhu, Lauga, and Brandt]{laipof1}
L.~Zhu, E.~Lauga, and L.~Brandt.
\newblock Self-propulsion in viscoelastic fluids: pushers vs. pullers.
\newblock \emph{Phys. Fluids}, 24:\penalty0 051902, 2012.

\bibitem[Keunings(1986)]{keunings1986}
R.~Keunings.
\newblock On the high weissenberg number problem.
\newblock \emph{J. Non-Newton. Fluid Mech.}, 30:\penalty0 209--226, 1986.

\bibitem[Owens and Phillips(2002)]{owens2002}
R.~G. Owens and T.~N. Phillips.
\newblock \emph{Computational rheology}, volume~14.
\newblock World Scientific, 2002.

\end{thebibliography}
\end{document}